\documentclass[journal=jpccck, manuscript=article]{achemso}
\setkeys{acs}{articletitle = true}
\usepackage{graphicx}
\usepackage{amsmath}
\usepackage{amssymb}
\usepackage{booktabs}
\usepackage{mhchem}

\title{Computational investigations of dispersion interactions between small molecules and graphene-like flakes}
\author{Tyler J.~Hughes}
\author{Robert A.~Shaw}
\author{Salvy P.~Russo} 
\email{salvy.russo@rmit.edu.au}
\affiliation{ARC Centre of Excellence in Exciton Science, School of Science, RMIT University, Melbourne, VIC 3000, Australia}

\begin{document}

\begin{tocentry}
\includegraphics[width=0.8\textwidth]{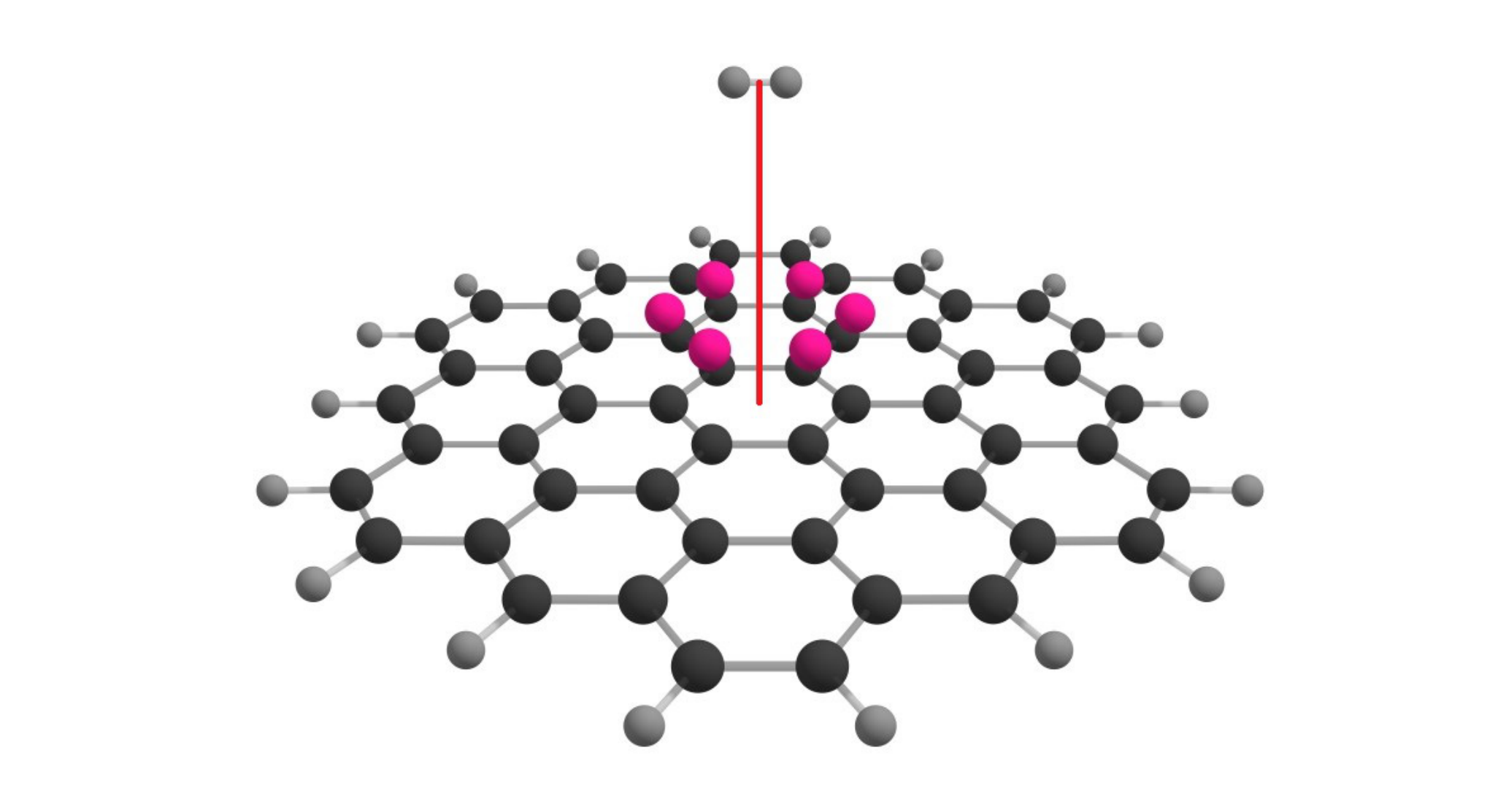}
\end{tocentry}

\begin{abstract}
In this work, we investigate dispersion interactions in a selection of atomic, molecular, and molecule-surface systems, comparing high-level correlated methods with empirically-corrected density functional theory (DFT). We assess the efficacy of functionals commonly used for surface-based calculations, with and without the D3 correction of Grimme. We find that the inclusion of the correction is essential to get meaningful results, but there is otherwise little to distinguish between the functionals. We also present coupled-cluster quality interaction curves for \ce{H2}, \ce{NO2}, \ce{H2O}, and \ce{Ar} interacting with large carbon flakes, acting as models for graphene surfaces, using novel absolutely localised molecular orbital based methods. These calculations demonstrate that the problems with empirically-corrected DFT when investigating dispersion appear to compound as the system size increases, with important implications for future computational studies of molecule-surface interactions.
\end{abstract}

\section{Introduction}

The development of new materials for molecular sensing is an important area of research, with applications ranging from the monitoring of harmful gas levels in the environment~\cite{Yoon2011, Lv2012, Chatterjee2015}, to the detection of explosives~\cite{Senesac2008, Germain2009, Hu2014}. The advantage such sensors would have over traditional spectroscopic methods is in their rapidity and ease of use. Additionally, in some cases, such materials could be used to selectively trap molecules, for example common pollutants~\cite{Kemp2013, Wang2013, Wang2015}. In order to be able to design materials with the desired properties, however, we need to understand how the molecules interact both with each other and the material surface. These interactions are characterised by a complex balance between different intermolecular forces, varying in strength and selectivity. 

In particular, many common gas molecules, such as hydrogen or nitrogen, have no permanent dipole or charge. As such, they interact predominantly through dispersion interactions, caused by the instantaneous responses of the electron distributions on each molecule to the presence of the other molecule. These are especially difficult to consider computationally, because they are purely quantum mechanical, and typically only constitute a small portion of the total energy. Thus, when taking energy differences to determine the interaction, any errors are significantly increased. Therefore, only the most accurate quantum mechanical approaches suffice to give a reliable description of these interactions~\cite{Rezac2011, Hohenstein2012}. 

When considering the large, periodic structures associated with materials, the difficulties are further amplified. High-level ab initio methods, such as coupled-cluster, cannot feasibly be applied - although periodic coupled-cluster methods are currently being developed~\cite{Hummel2017, Gruber2018, Jordan2019}. The most common choice of method is then density functional theory (DFT). In principle, by choosing a suitable functional, it could be possible to get very accurate results, and the extension to periodicity is reasonably simple~\cite{Hasnip2014}. However, there is an inherent uncertainty involved in choosing a functional, which depends on the design and parametrisation of the functional, and often depending on the exact system being studied~\cite{Zhao2005, Marom2011, Taylor2016}. This makes it difficult to consistently get reliable results. 

Many previous studies have looked at benchmarking the gas-phase interactions of small molecules with each other, by comparing high-level results with various density functionals~\cite{Kozuch2013, Rezac2015, Rezac2016, Claudot2018, Shaw2019}. More complex functionals, such as double hybrids with some element of ab initio correlation energy built in, sometimes perform quite well~\cite{Kalai2019}. They are considerably more expensive than single hybrids, however, and often perform very poorly for pure dispersion interactions~\cite{Goerigk2011}, while showing a high dependence on choice of basis set~\cite{Karton2011}. Instead, it is more typical to either directly incorporate interactions into the parametrisation of the functional, such as with the Minnesota functionals~\cite{Zhao2008, Zhao2011}, or add-on an empirical dispersion correction~\cite{Tkatchenko2009, Grimme2011}. The most popular of these is the D3 correction of Grimme~\cite{Grimme2010}, which has been repeatedly shown to give reasonable results across several different benchmark sets~\cite{Goerigk2011, Goerigk2011b, Schroder2017}. 

The explicit benchmarking of these corrected functionals on larger molecular interactions, as would be found in molecule-surface interactions, is less well explored~\cite{Reckien2014, Rehak2020}. This is unsurprising, given the aforementioned difficulty in obtaining high level results. However, recent developments in the area of large-scale coupled cluster calculations - in particular, localised orbital methods~\cite{Ma2019, Shaw2019b} - now opens the possibility of determining which functionals can most reliably give accurate results for dispersion interactions between molecules and surfaces. In the present paper, we explore various functionals, with and without empirical corrections, for just this purpose. We begin by considering a series of predominantly dispersive small molecular interactions, allowing us to determine both which correction to use and the performance of our chosen high-level method. This is then followed by investigating the interaction of a series of small molecular gases with graphene-like carbon flakes of increasing sizes. The recommendations drawn from this should then provide confidence in future studies of such interactions in extended systems.

\section{Computational methodology}

The systems considered herein are divided into three categories: pairs of noble gases, \ce{Ne2}, \ce{Ar2}, \ce{Kr2}, \ce{Ne-Ar}, and \ce{Ar-Kr}; small molecules,en comprising dimers of \ce{H2}, \ce{CH4}, ethyne, benzene and pyridine, plus benzene and pyridine interacting with \ce{H2}, \ce{CH4}, ethyne, \ce{H2O}, and \ce{MeOH}; and molecule-surface models, with \ce{H2}, \ce{NO2}, \ce{H2O}, and \ce{Ar} above hydrogenated carbon flakes of various sizes, as will be discussed later. These systems were chosen as they are representative of dispersion-dominated interactions. The possible exceptions are those involving \ce{NO2} and \ce{H2O}, which have dipole moments, such that other effects may be important. They are, however, very important in sensing applications~\cite{Chung2012}, and water is one of the only molecules for which there are previous high-level results to compare to~\cite{Voloshina2011,Brandenburg2019}. Optimised equilibrium geometries for all these systems can be found in the Supporting Information. 

\begin{figure}

\begin{tabular}{cccc}
\includegraphics[width=0.24\textwidth]{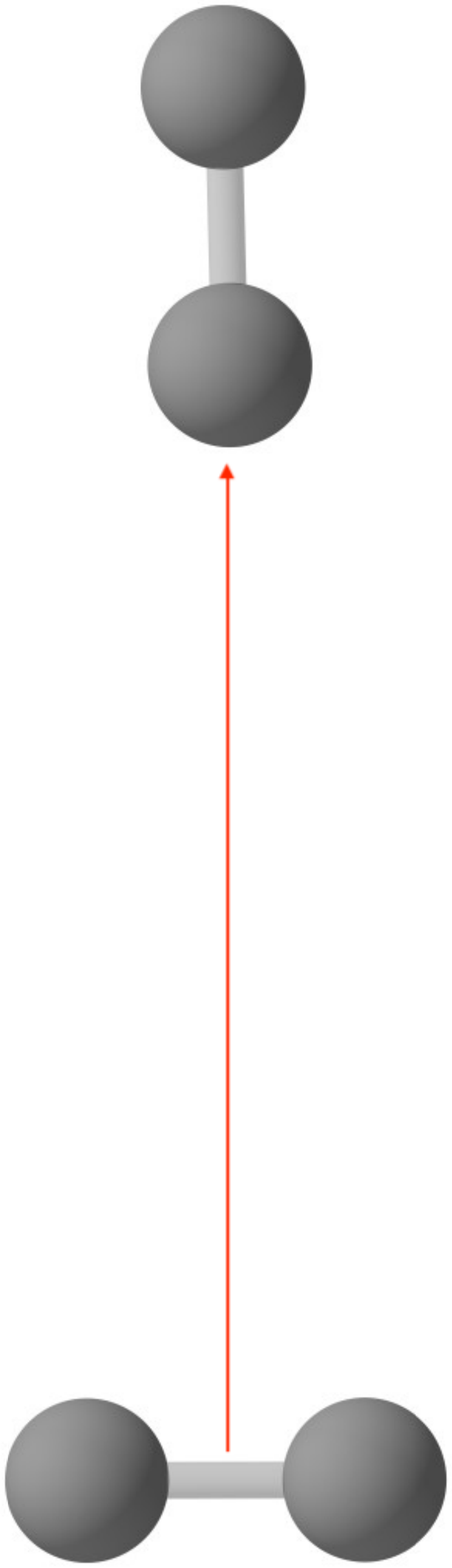} &
\includegraphics[width=0.24\textwidth]{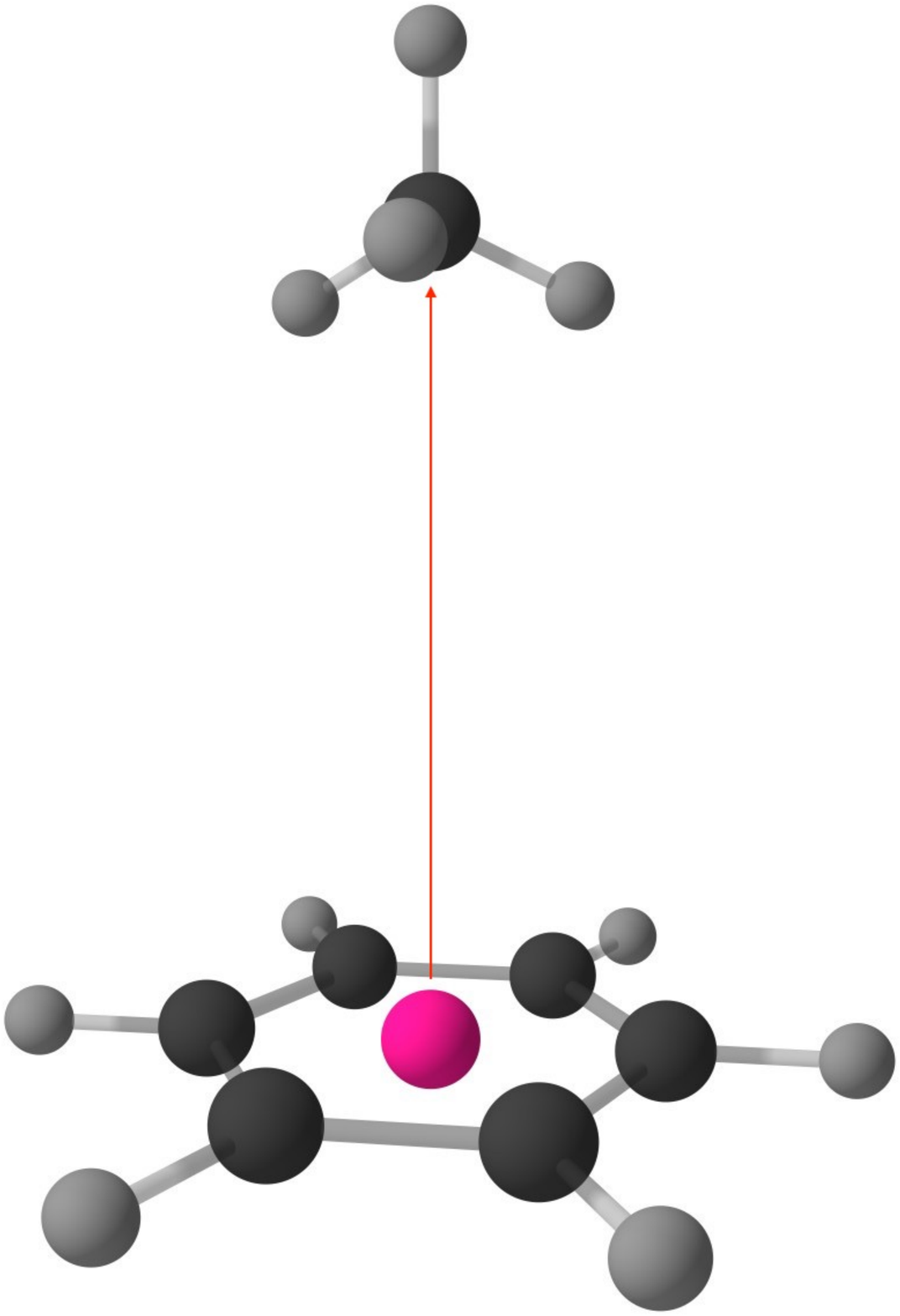} &
\includegraphics[width=0.24\textwidth]{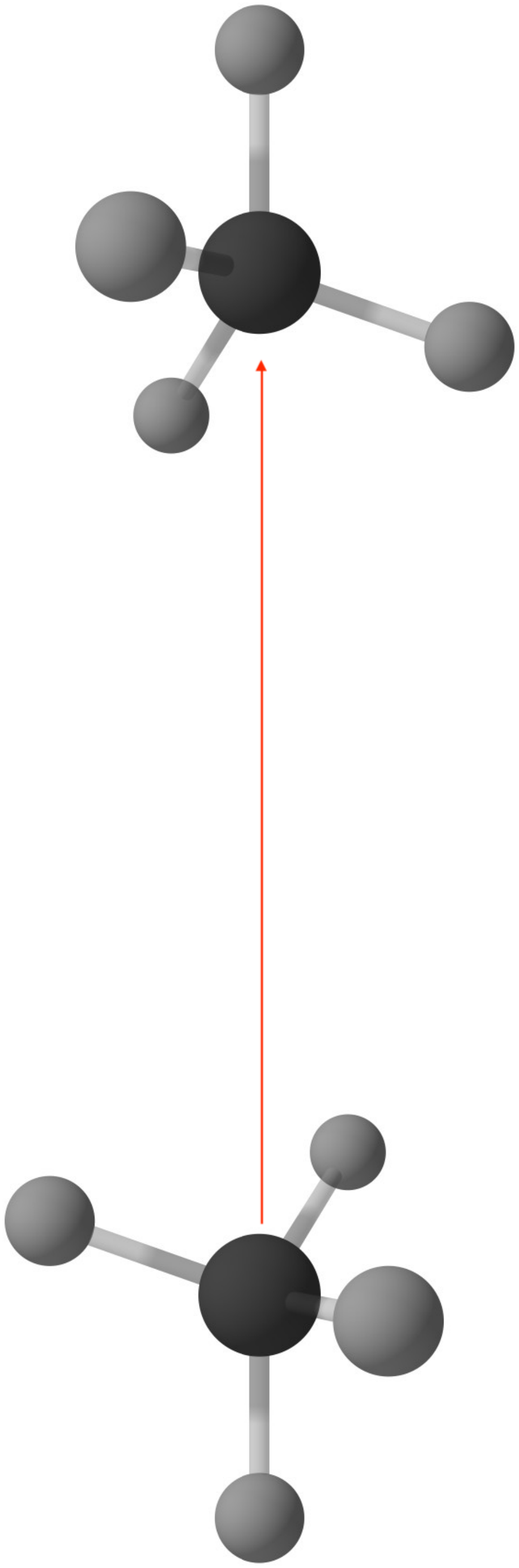} &
\includegraphics[width=0.24\textwidth]{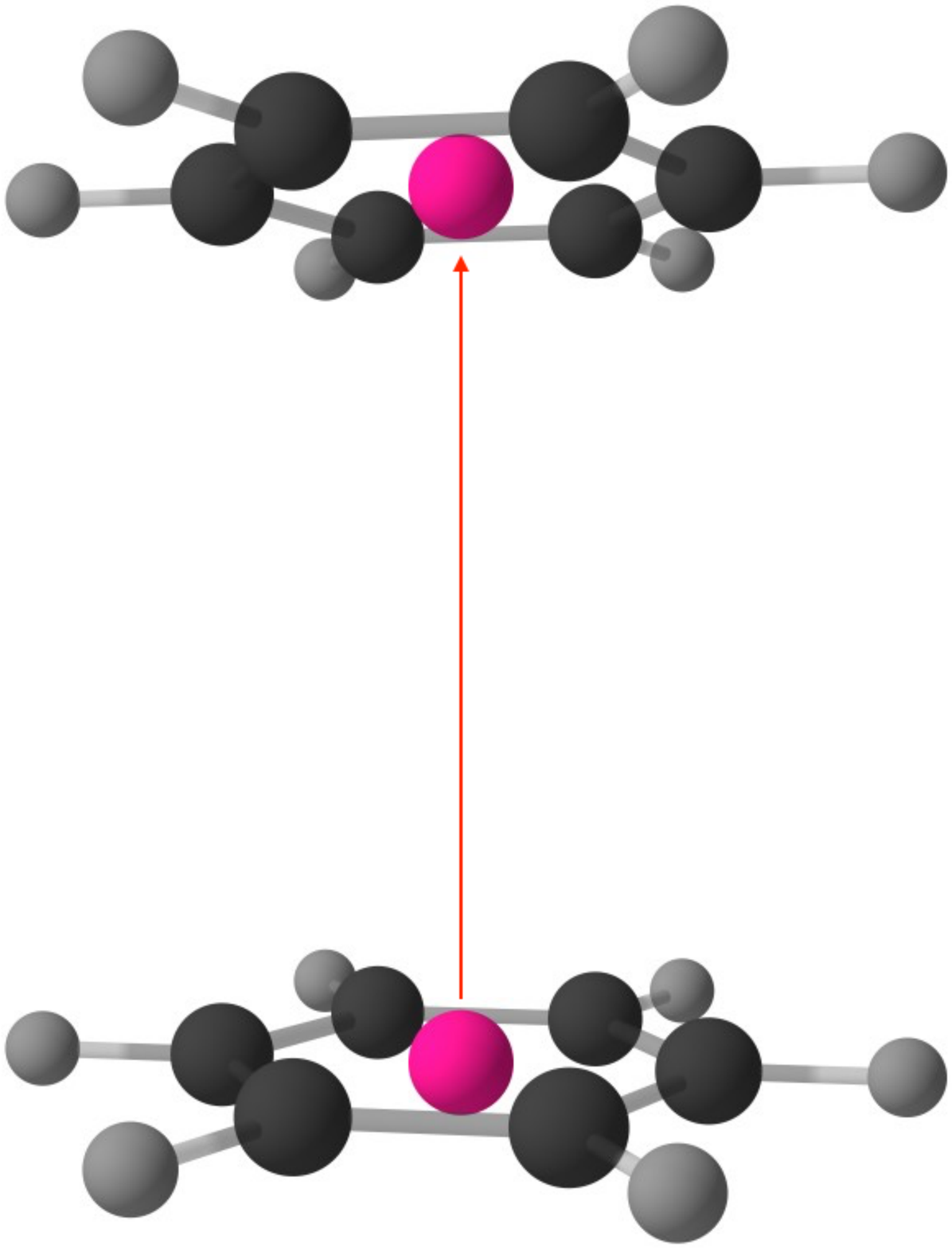}   
\end{tabular}

\caption{Structures of some of the molecular systems, with the intermolecular separation, $R$, marked as a red arrow. From left to right, these are \ce{(H2)2} (t-shaped), benzene-methane, methane dimer, and benzene dimer (pi-pi stacked). Pink `ghost' atoms are only there to mark where the central points are, to make the definition of $R$ clearer.}
\label{fig.molecules}
\end{figure}

For the small molecule interactions, initial geometries were optimised using coupled-cluster with singles and doubles excitations and perturbative triples [CCSD(T)]. For each of these, and the noble gases, interaction energy curves were determined by varying the intermolecular separation from 2.5 to 5~Angstroms, without relaxing the geometry at each step. Examples of the chosen axis for each molecular case are shown in Figure~\ref{fig.molecules}. These scans were performed at the CCSD(T) level, and replicated with the PBE~\cite{Perdew1992}, TPSS~\cite{Tao2003}, PBE0~\cite{Perdew1996}, B3LYP~\cite{Becke1988, Lee1988}, and B98~\cite{Schmider1998} functionals, each with and without the Grimme D3 correction. Wherever this D3 correction is used, we also apply Becke-Johnson (BJ) damping~\cite{Grimme2010, Grimme2011}, but we suppress the (BJ) to shorten already overlong functional acronyms. In addition, we have included the M06-2X~\cite{Zhao2008} and recent $\omega$B97M-V functionals, without any corrections, as examples of functionals where dispersion interactions are a part of the design and parameterisation of the functional itself. The basis set used on all atoms was aug-cc-pVTZ in every instance~\cite{Kendall1992}. All DFT and CCSD(T) calculations were performed in the \textsc{ORCA} suite of programs~\cite{Neese2012}. All CC and DFT interaction energies have been calculated with the counterpoise correction of Boys and Bernardi~\cite{Boys1970}, to eliminate basis set superposition errors.

In addition, the same points were calculated using the absolutely localised molecular orbital (ALMO) method~\cite{Stoll1980, Khaliullin2006} with a local ring coupled cluster approximation (rCCD, or RPAx)~\cite{Shaw2019b}. These used density fitting on the coulomb-exchange integrals and the integral transformation, with the AVTZ/JKFit and MP2Fit auxiliary basis sets for all atoms~\cite{Weigend2002, Weigend2002b, Weigend2006}. The local Fock build, infinite-order charge transfer correction, and second-order screened exchange correction, were used in each case, along with perturbative triples calculated as per CCSD(T)~\cite{Deegan1994}, but with the rCCD amplitudes. For convenience, we term the overall method ALMO+rCCD(T), and has been implemented in the \textsc{Gamma} software~\cite{Gamma2020}. 

\begin{figure}

\begin{tabular}{cc}
\includegraphics[width=0.45\textwidth]{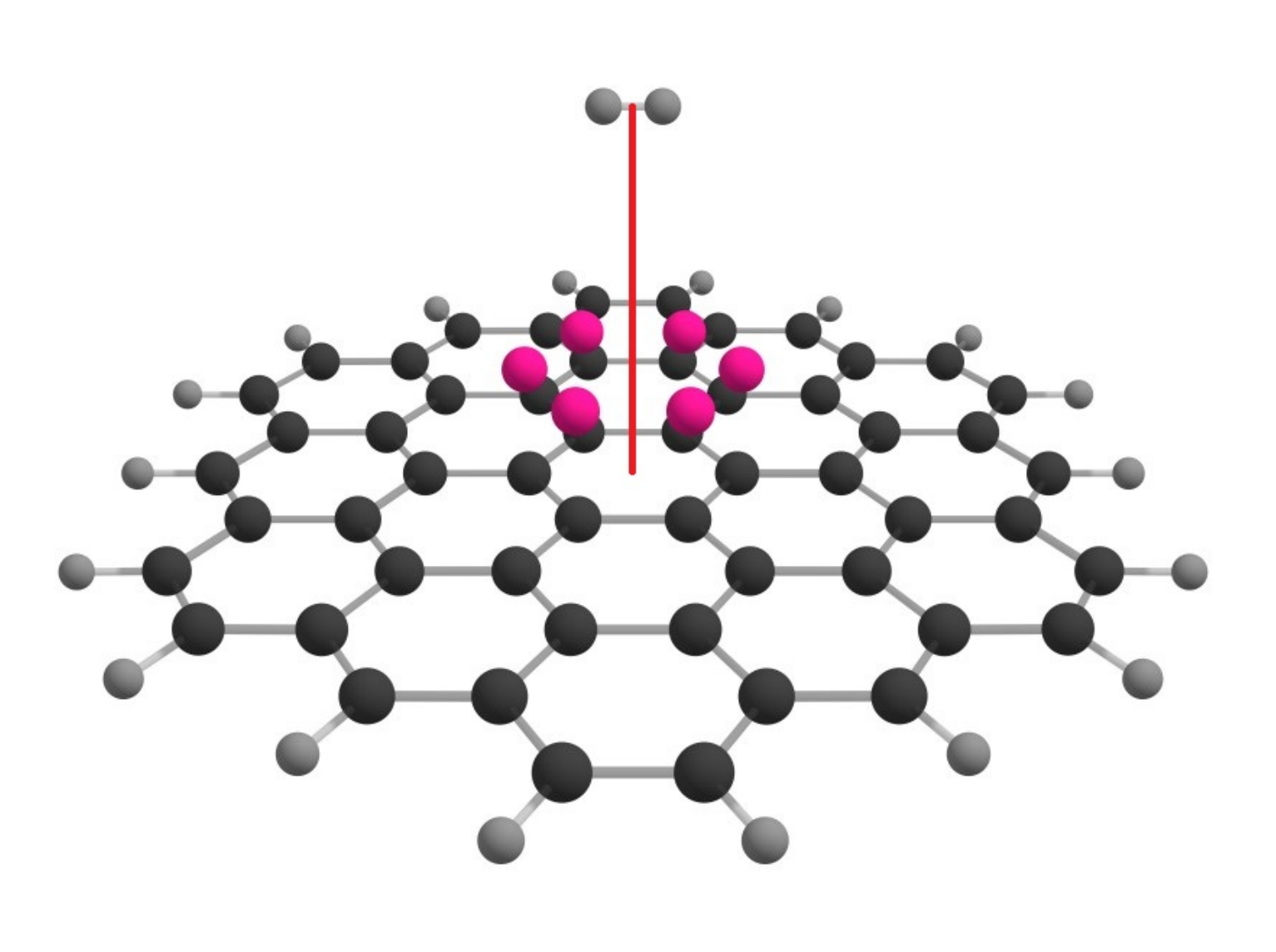} &
\includegraphics[width=0.45\textwidth]{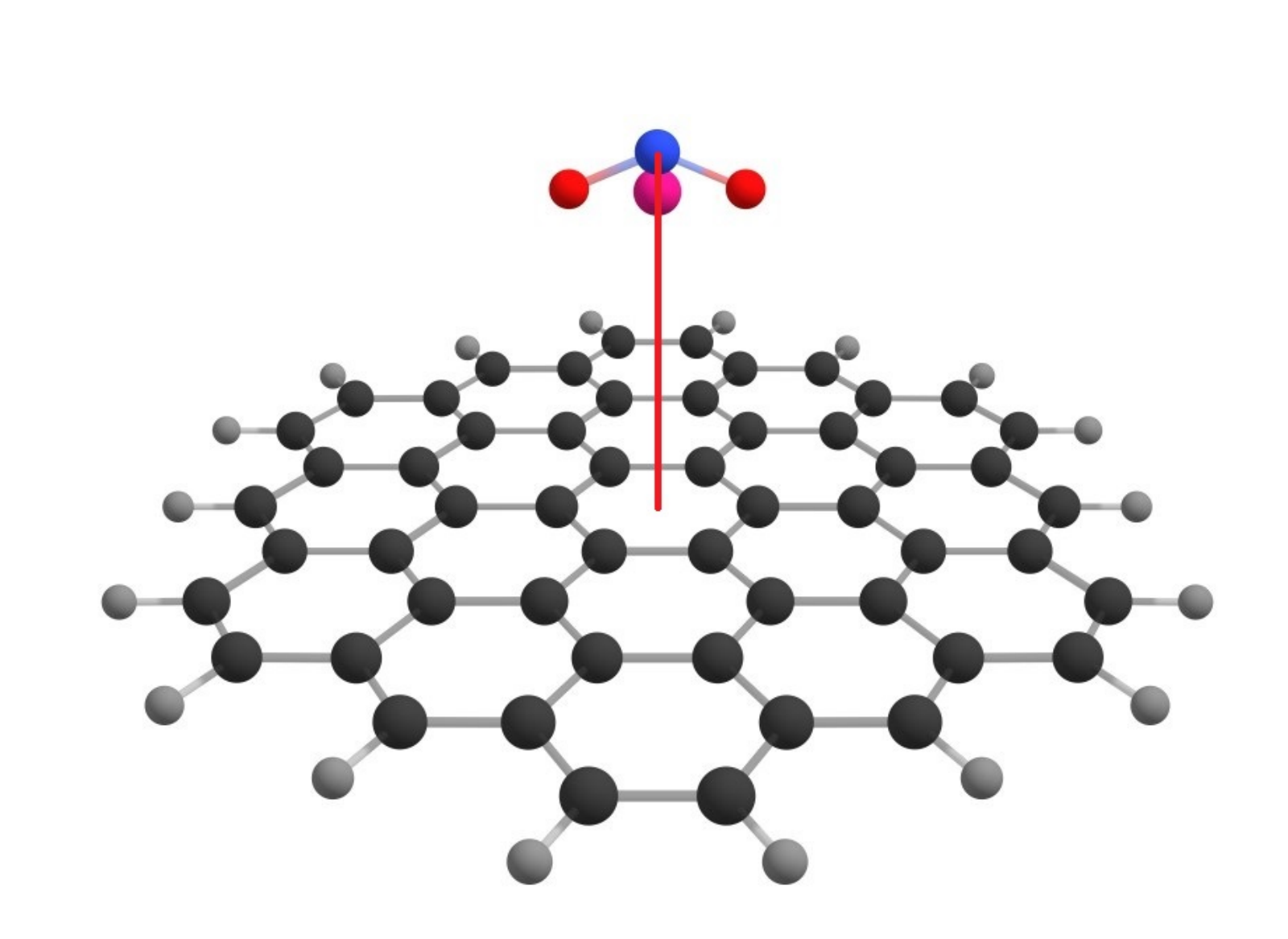} 
\end{tabular}

\caption{Structures of the largest carbon flakes - circumcoronene - interacting with hydrogen (left) and \ce{NO2} (right). The orientations for the smaller flakes are the same, but with a ring of hexagons removed in each case. Pink atoms are ghost atoms, with a hydrogen basis for \ce{H2} and helium basis for \ce{NO2}. The structures with \ce{H2O} and \ce{Ar} follow those with \ce{NO2}, with the ghost atom in the same location.}
\label{fig.surfaces}
\end{figure}

For the molecule-surface models, initial geometries were optimised with B3LYP+D3 and an AVTZ basis, with the resolution of the identity approximation (RIJK) used on both the Coulomb and exchange integrals. The AVTZ/JKFit auxiliary bases were used~\cite{Weigend2002, Weigend2002b, Weigend2006}.  The molecular placements for the largest flakes, along with ghost atoms (described shortly) are shown in Figure~\ref{fig.surfaces}. Scans were performed on the separation between the molecule and that plane of the surface in the same manner as for the smaller systems, but using only the B3LYP+D3 and ALMO+rCCD(T) methods.

In the hydrogen molecule systems, ghost atoms with the AVTZ hydrogen atom basis were placed an Angstrom above each of the 6 carbon atoms closest to the \ce{H2}. With \ce{NO2}, \ce{H2O}, and \ce{Ar}, we instead placed a single ghost atom, with an AVTZ helium basis, 2~\AA~above the plane of the surface but directly below the central atom. These were found to be necessary to be able to achieve convergence in the Hartree-Fock procedure at longer separations whenever the carbon flake was bigger than benzene, and aided convergence even in that case. The difficulty converging is predominantly due to the diffuse functions on the passivated hydrogens approaching linear dependency. This could be solved by removing the most diffuse exponents from the hydrogens, but this then affects the quality of the interaction energy, where diffuse functions play an important role. Inclusion of ghost atoms alleviates this issue by greatly improving the description of the region between the monomers, reducing the importance of the diffuse hydrogen functions in the occupied orbitals, thus stabilising the convergence of the Hartree-Fock density.

\section{Results and discussion}

We firstly need to compare the efficacy of a selection of density functionals for the types of systems that we are interested in. Ultimately, the aim is to run calculations on extended molecule-graphene systems, which we anticipate will be predominantly dispersive in nature. As mentioned earlier, there are two main routes to including dispersion in DFT calculations: via parametrisation into the functional itself, or via empirical dispersion corrections. There is also a third possibility of including some component of correlation through RPA- or MP2-like additions~\cite{Kozuch2011, Chermak2012}; these are, however, relatively expensive and thus not appropriate for our considerations. Our focus here is on the empirically-corrected functionals, using Grimme's D3 correction, because these are computationally the cheapest, with the correction adding essentially no cost while often being very effective. To this end, we have chosen five functionals - a GGA (PBE), a meta-GGA (TPSS), and three hybrids (PBE0, B3LYP, and B98), all of which are commonly used in periodic calculations on the solid state. We have additionally included M06-2X and $\omega$B97M-V, as examples of uncorrected functionals that are known to perform well for intermolecular interactions; they are not typically used for extended systems, however. We will compare the counterpoise-corrected interaction energies calculated with these, with CCSD(T) results, for a range of noble gas dimers and small molecular systems, all of which are bound mainly through dispersion forces.

\begin{figure}

\begin{tabular}{cc}
\includegraphics[width=0.49\textwidth]{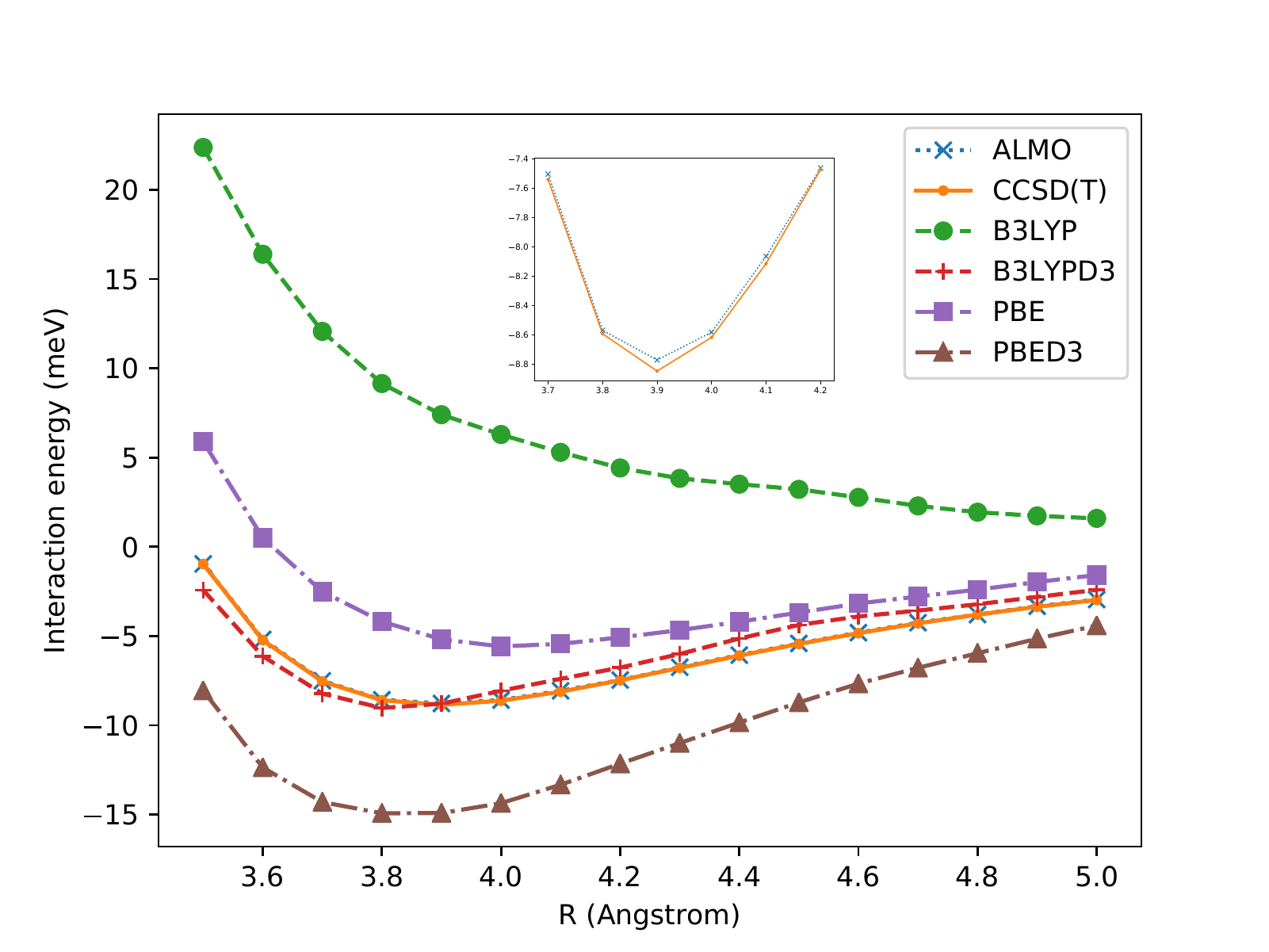} &
\includegraphics[width=0.49\textwidth]{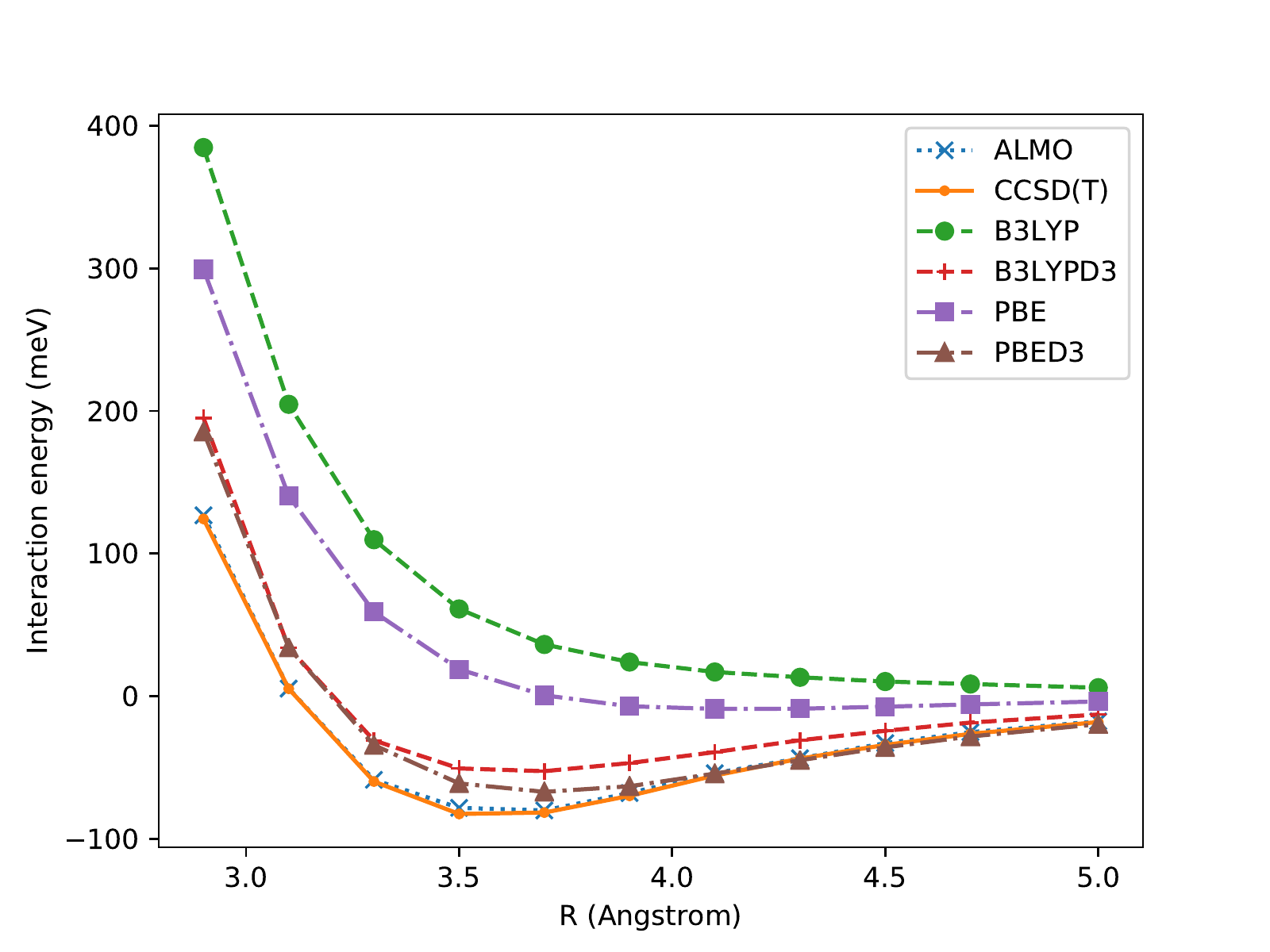} \\
\ce{Ar2} & Benzene\ce{\bond{-}CH4}
\end{tabular}

\caption{Interaction energy curves for the argon dimer (left) and benzene-methane complex (right), for a selection of different methods. The ALMO+rCCD(T) and CCSD(T) curves appear indistinguishable, so the inset for the argon dimer shows a zoomed in portion around the minimum for just these two.  In both systems, addition of the D3 correction to the density functionals is essential to give reasonable results.}
\label{fig.curves}
\end{figure}

Figure~\ref{fig.curves} shows curves for two representative systems in each class: the argon dimer and benzene-methane complex. For clarity, we have only shown the corrected and uncorrected PBE and B3LYP functionals; all other results can be found in the Supporting Information, and summary statistics will be discussed later. From the figure, it is clear that without the D3 correction, neither functional performs well for either type of system. In fact, both PBE and B3LYP give dissociative curves for the benzene-methane complex, while only PBE shows any binding for the argon dimer. This is borne out with the other functionals too, with TPSS and PBE0 following the same trends as PBE, and B98 following B3LYP. Adding the D3 correction greatly improves the shape of the curves, appearing to give better agreement with CCSD(T) on the equilibrium geometry. However, for the noble gases where all but B3LYP and B98 were already giving some dispersion, this results in a significant overcorrection, and thus overbinding of the complex, by roughly 200\% in the case of PBE+D3. It should be noted that both PBE and PBE0 were created with reference to RPA correlation energies~\cite{Perdew1992, Perdew1996}, which give the exact long-range $R^{-6}$ dependence of dispersion, so it is possible this overcorrection is effectively double counting the dispersion contribution. However, the D3 correction was parametrised for each functional separately, so we would not expect this to be the case more generally. 

Also shown in Figure~\ref{fig.curves} are the ALMO+rCCD(T) results for the same systems, which appear to be effectively indistinguishable from the CCSD(T) results. The reason for this is that the errors are on the order of 2~meV (0.2~kJmol$^{-1}$), as can be seen in the inset on the left-hand panel of the figure. This is in comparison to even the best DFT results, which are in error on the order of 10~meV. Incidentally for these systems, the ALMO calculations are actually less computationally expensive, at roughly half the wall time on average, than even the cheapest density functional. These speedups will not, however, hold for much larger systems, as the method was designed for multiple small fragments rather than one extended system~\cite{Shaw2019b}. The results are promising for the use of ALMOs as a comparison in the midrange between molecular and extended systems, as will be discussed later when considering carbon flakes. 

\begin{figure}

\includegraphics[width=0.7\textwidth]{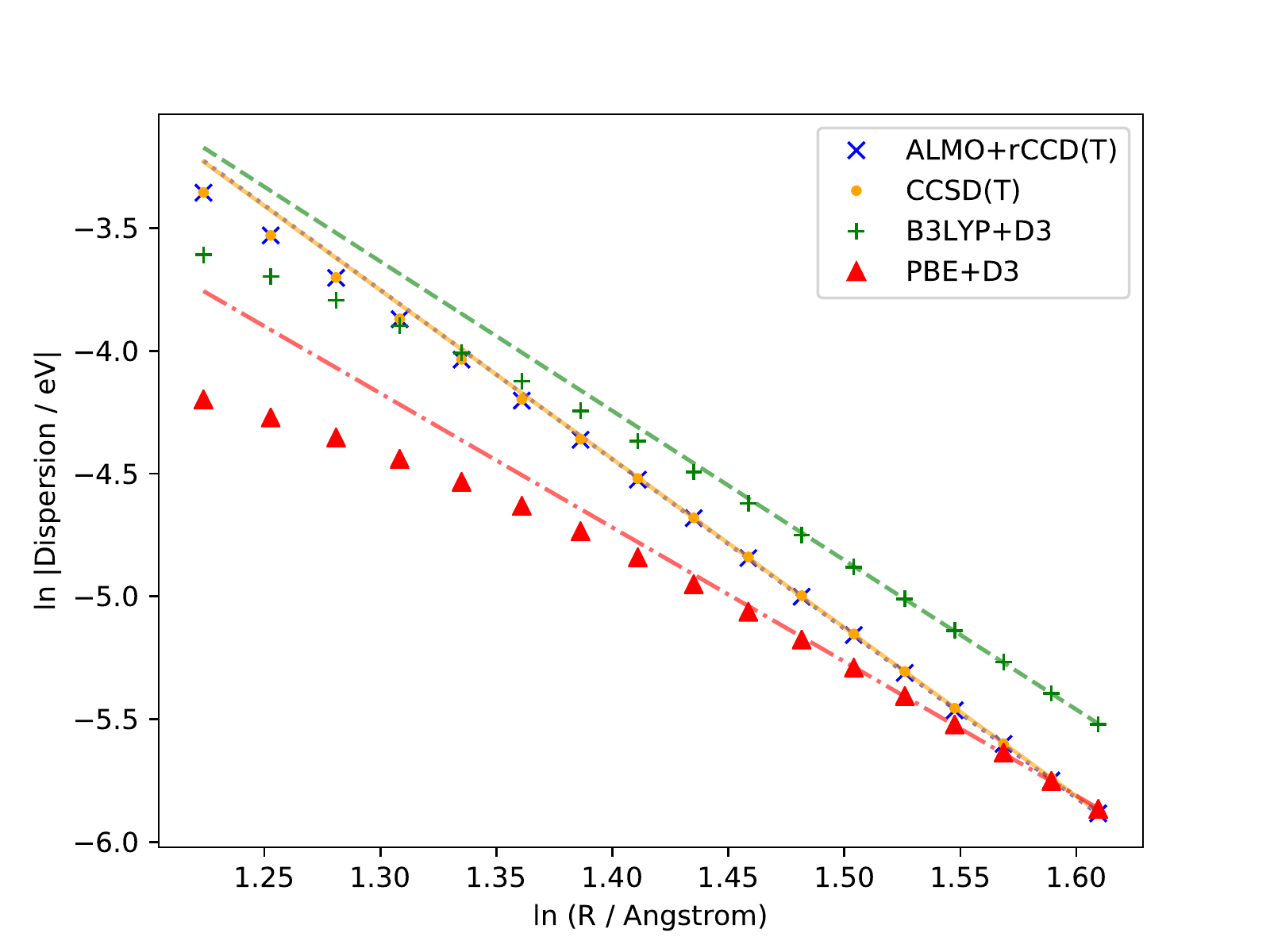}

\caption{A log-log plot of the dispersion contribution for a selection of methods, with straight lines fitted to the points from 4.5 to 5~\AA~($\ln(R)=1.5$ to 1.6). For ALMO+rCCD(T) and CCSD(T), which again largely overlap, the dispersion is determined as the correlation energy contribution, while for the density functionals, it is the D3 correction.}
\label{fig.dispersion}
\end{figure}

We can see in more detail what is happening with the dispersion energies for the argon dimer in Figure~\ref{fig.dispersion}, which shows just the D3 correction terms compared with the dispersion contributions from CCSD(T) and ALMO+rCCD(T) (in this case, this is equivalent to the correlation energy, as this is a neutral diatomic system). The long-range behaviour should follow a strict $R^{-6}$ dependence, resulting in a linear log-log plot where the intercept of the line gives the dispersion coefficient. From the figure, it's clear that the D3 correction is behaving very differently from the wavefunction-theory correlation energies. While the latter approach straight lines with equivalent slopes and intercepts at around 4.1~\AA~($\ln(R)=1.4$), the D3 results are much more curved and may not yet be linear at any point within the distances considered (up to 5~\AA). This is a demonstration of how the correction is having to compensate for an uneven inclusion of correlation energy in the original functionals, with the anisotropy most apparent for the PBE-based correction (the PBE0 and TPSS corrections are similar). This suggests that B3LYP, by essentially including minimal dispersion in its original formulation, is best placed of those selected to add it through an ad hoc correction. 

It is worth noting at this point that the $C_6$ dispersion coefficients that could be obtained from Figure~\ref{fig.dispersion} are not worth considering, as it is well-known that noble gas interactions are very long range~\cite{Cybulski1992}, and contain a considerable amount of correlation energy due to core electrons~\cite{Ranasinghe2015}. This means that correlated calculations should really use core-valence basis sets with the core electrons unfrozen in the correlated part of the calculation. For the heavier elements, in particular krypton of those considered here, there are also relativistic effects that may need to be taken into account~\cite{Nicklass1995}. The long range of these interactions also warrants the use of ghost atoms, which can have a drastic effect on both the equilibrium separation and interaction energy~\cite{Patkowski2013, Shaw2018}. It is not reasonably possible to account for such effects in the DFT calculations - where no special consideration is given to the core electrons and basis set dependence is much lower - which is why we have not included them here. 

\begin{figure}

\includegraphics[width=0.8\textwidth]{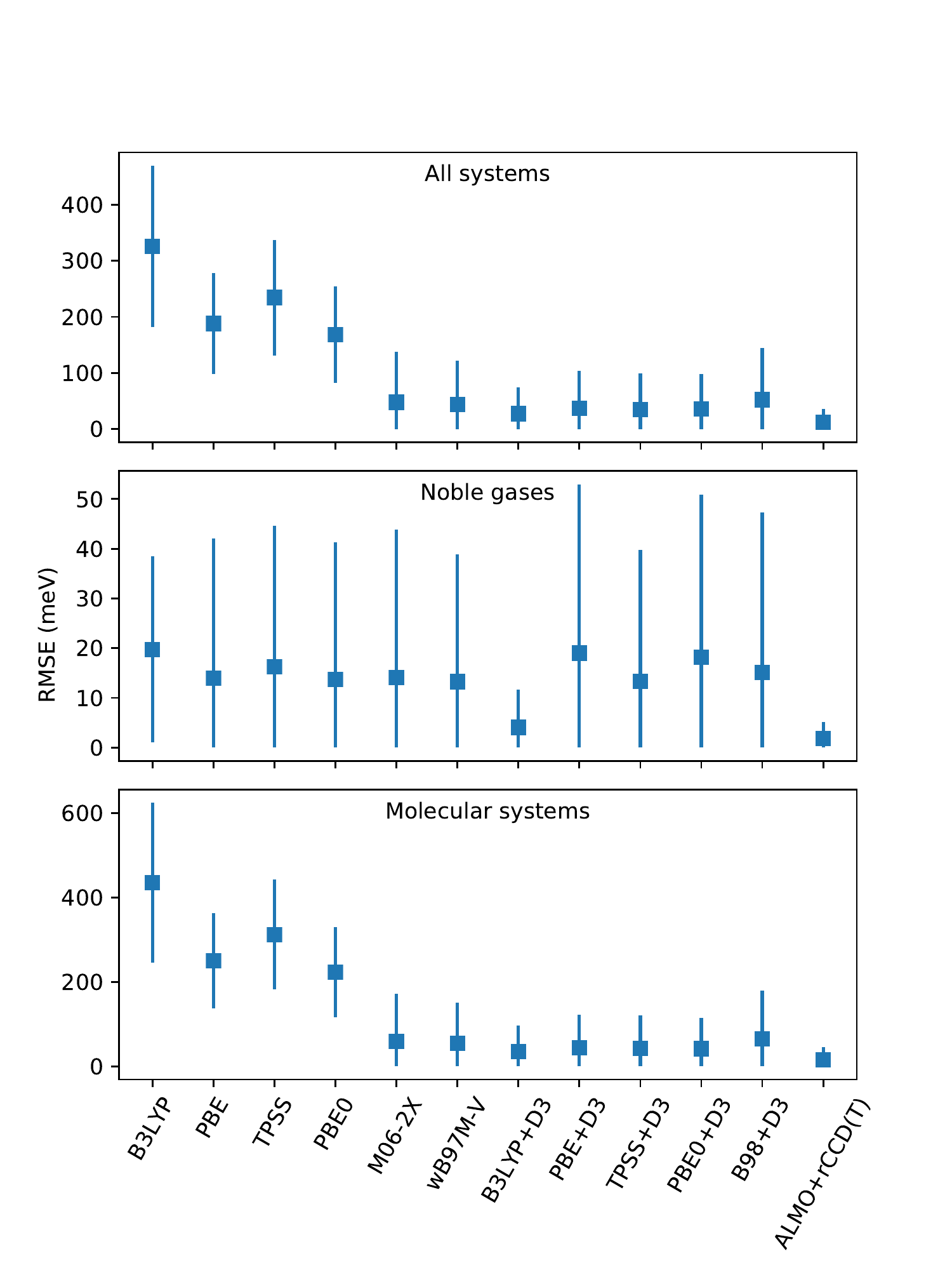} 

\caption{Error distributions relative to CCSD(T)/AVTZ results for all methods used, split by type of system. The root-mean-square error across all geometries are the solid squares, with error bars showing 1.95 standard deviations of the distributions.}
\label{fig.errors}
\end{figure}

The overall trends noted above seem to repeat across all the systems we investigated, as demonstrated in Figure~\ref{fig.errors}. Here we can see the root-mean-square errors in the interaction energies for each method, compared to CCSD(T), for all geometries. Firstly, it is clear that ALMO+rCCD(T) is giving results essentially equivalent to CCSD(T) when considered relative to the errors from using DFT. The overall RMSE across all systems is 11.9~meV, with a standard deviation of roughly 10~meV. This is within chemical accuracy, and moreover is consistent across all the systems, with the largest error (44~meVl) being for the benzene-water interaction. The perturbative triples contribution makes a substantial difference in the molecular systems, reflecting the importance of moving beyond pairwise approximations when calculating dispersion. We would expect second-order perturbation theory to perform fairly poorly, if that were to be used as an alternative `cheap' correlated method. 

For the density functionals, there is a clear trend of the uncorrected functionals performing very poorly regardless of type, with the addition of the D3 correction reducing the error by an order of magnitude. This is most pronounced for the molecular systems, where the B3LYP error is around 435~meV, whereas B3LYP with the correction is around 35~meV. Interestingly, for the noble gases, the D3 correction actually increases the error for some systems when using the PBE, PBE0, and TPSS functionals. This is a continuation of the overestimation of dispersion found in Figure~\ref{fig.curves}, and is predominantly just for two of the systems (the argon dimer and argon-krypton complex). It is possible that the interaction energies for these systems are so small that any small error presents as a large percentage error.  This reflects in the fact that the RMSE and spread for the D3-corrected functionals is largely the same for both the molecular and noble gas systems, they just appear bigger due to the scale. Overall, a mean error of 40~meV (4~kJmol$^{-1}$) is reasonable, but on the boundary of what is considered acceptable chemical error. 

\begin{figure}

\begin{tabular}{cc}
\includegraphics[width=0.49\textwidth]{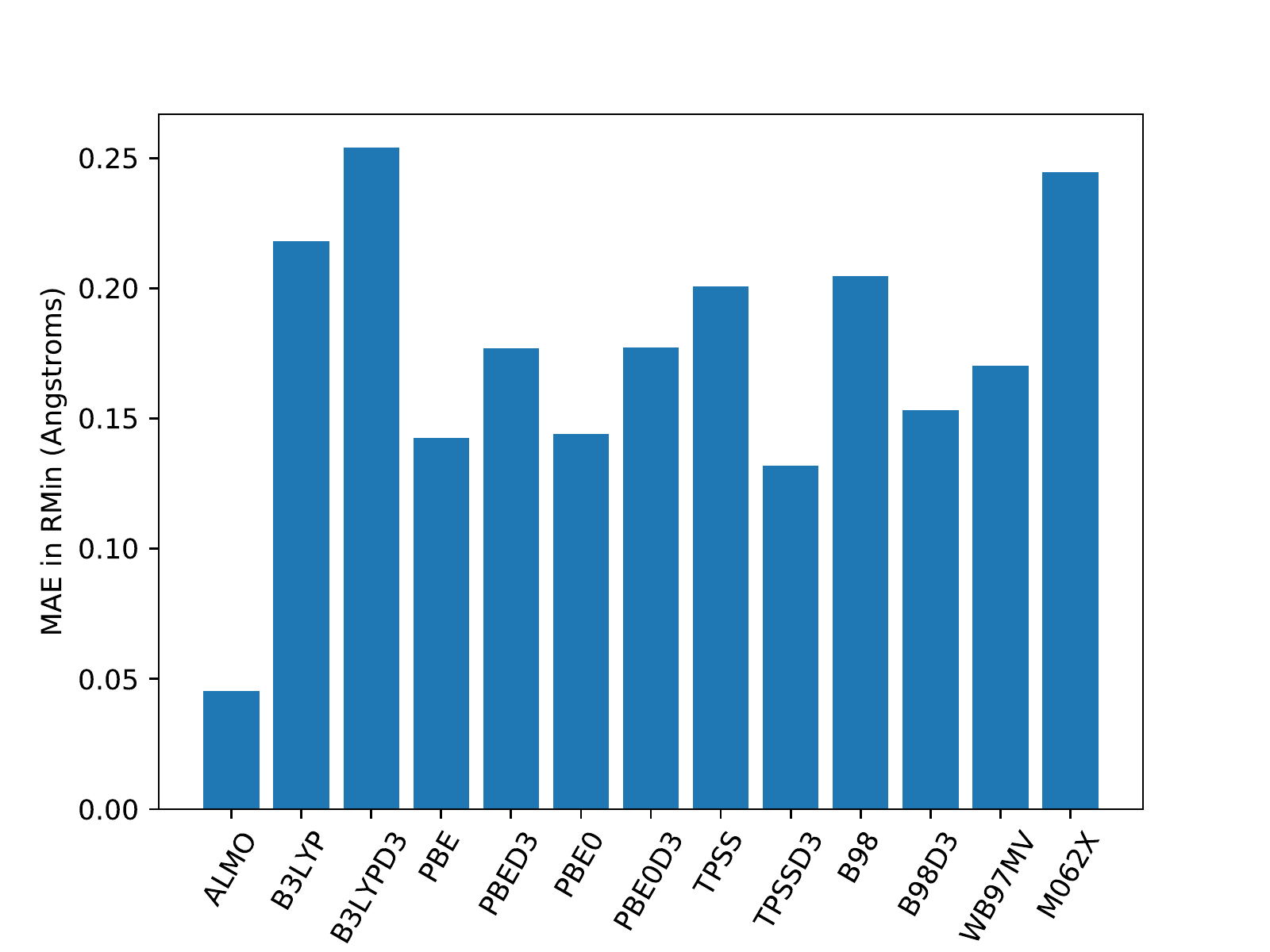} &
\includegraphics[width=0.49\textwidth]{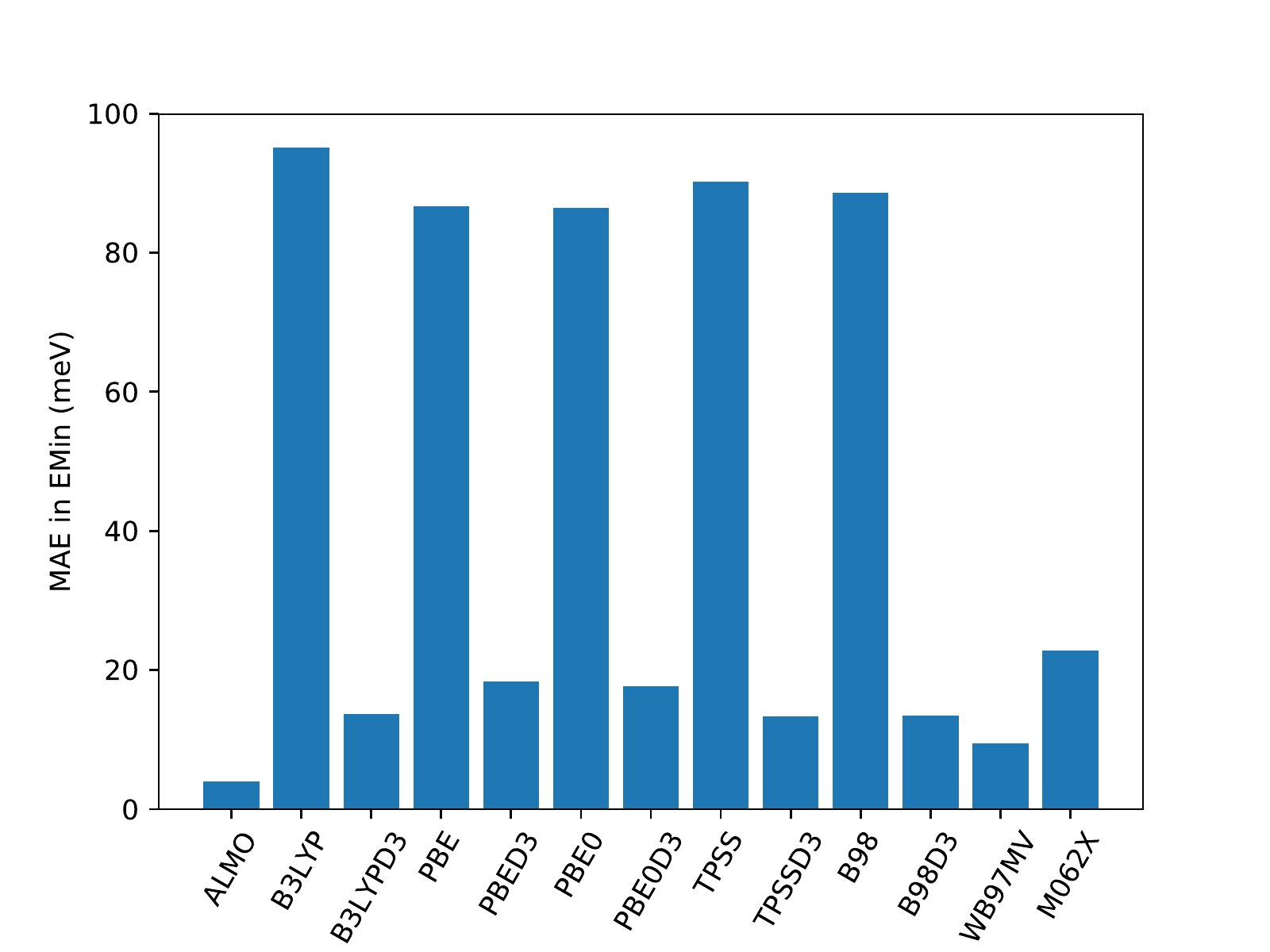} 
\end{tabular}

\caption{Mean absolute errors in the binding constants across all systems considered, for each method, relative to the CCSD(T)/AVTZ results. On the left is the error in the equilibrium separation, while on the right is the error in well depth.}
\label{fig.properties}
\end{figure}

Equally important as the interaction energy, however, are the binding properties, or so-called binding constants. Specifically, we look at the equilibrium separation, $R_0$, and well-depth, $\epsilon_0$. These were calculated by fitting a quadratic to a five-point stencil centred on the lowest calculated energy of each system for each method. These properties for a couple of representative systems can be found in Table~\ref{table.specconsts}, while results for all systems considered can be found in the Supporting Information. Figure~\ref{fig.properties} gives summaries of the overall performance of each method.  Again, we see that the functionals all perform similarly, but we note that for the predicted binding distance, only B98 and TPSS perform better upon addition of the D3 correction. All the corrected functionals show a large improvement in predicted $\epsilon_0$. Overall, the combination of these results and those from Figure~\ref{fig.errors}, plus the problems with noble gases for other functionals, makes us select B3LYP+D3 as the functional to use for the larger systems in the next section. It is, however, somewhat arbitrary, and we would make no strong recommendation other than that the use of the correction is essential, as has been well documented elsewhere~\cite{Schroder2017}. 

\begin{table}
\caption{Binding constants for the argon dimer and benzene-methane complex. Results for all other systems can be found in the Supporting Information. The missing data correspond to dissociative interaction curves.}
\label{table.specconsts}

\begin{tabular}{lcccc}
\toprule
& \multicolumn{2}{c}{\ce{Ar2}} & \multicolumn{2}{c}{Benzene\ce{\bond{-}CH4}} \\
Method & $R_0$ (\AA) & $\epsilon_0$ (meV) & $R_0$ (\AA) & $\epsilon_0$ (meV) \\
\midrule
     CCSD(T) &  4.00 & -9.29 &  3.59 & -106.71 \\
        ALMO+rCCD(T) &  4.00 & -9.23 &  3.59 & -103.80 \\
       B3LYP &  -- &  -- &  --  &  -- \\
     B3LYP+D3 &  3.96 & -8.72 &  3.66 & -59.61 \\
         PBE &  4.06 & -6.35 &  3.79 & -7.71 \\
       PBE+D3 &  3.93 & -15.17 &  3.69 & -72.97 \\
        PBE0 &  4.08 & -4.34 &  3.79 & -4.83 \\
      PBE0+D3 &  3.96 & -13.36 &  3.68 & -67.68 \\
        TPSS &  4.16 & -2.94 &  -- &  --  \\
      TPSS+D3 &  4.03 & -14.08 &  3.70 & -68.80 \\
         B98 &   -- &  -- &  -- &  -- \\
       B98+D3 &  4.07 & -13.02 &  3.67 & -85.71 \\
      $\omega$B97M-V &  3.95 & -10.92 &  3.67 & -58.72 \\
       M06-2X &  4.07 & -8.86 &  3.46 & -125.23 \\
\bottomrule
\end{tabular}

\end{table}

Interestingly, neither of the more `modern' functionals, M06-2X or $\omega$B97M-V, perform any better than the D3-corrected hybrids (B98 and B3LYP). The Minnesota functional is in some cases considerably worse, with an overall RMSE for the molecular systems of 60~meV, roughly twice that of B3LYP+D3. Overall, $\omega$B97M-V probably gives the best results across all the functionals considered, with total errors in the interaction energies largely the same as for the best corrected hybrid, but somewhat better prediction of the binding constants. However, as there is negligible difference and it is not currently widely available in quantum chemistry codes, we have chosen to use B3LYP+D3 for subsequent investigations.

\begin{figure}

\begin{tabular}{cc}
\includegraphics[width=0.49\textwidth]{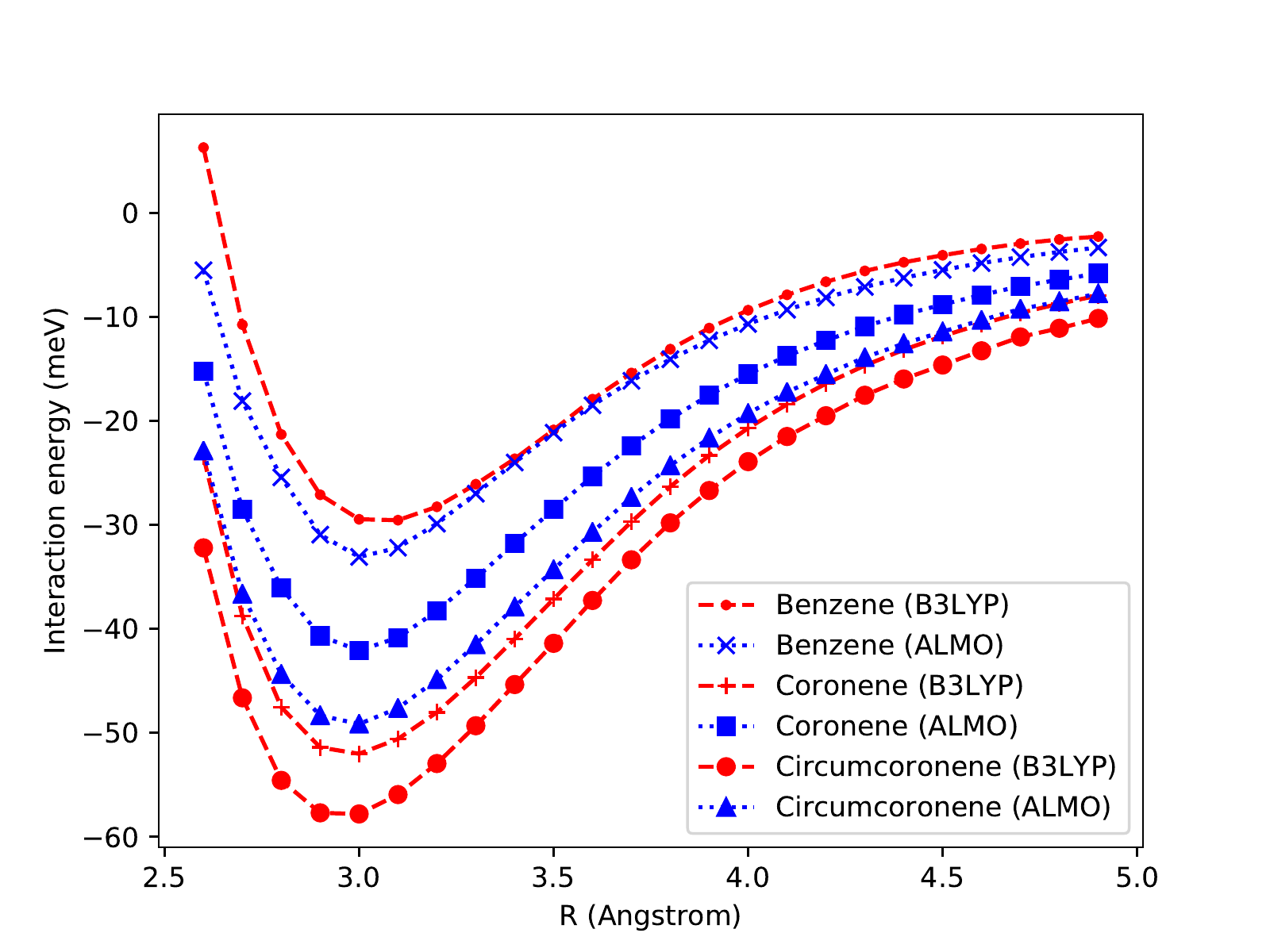} &
\includegraphics[width=0.49\textwidth]{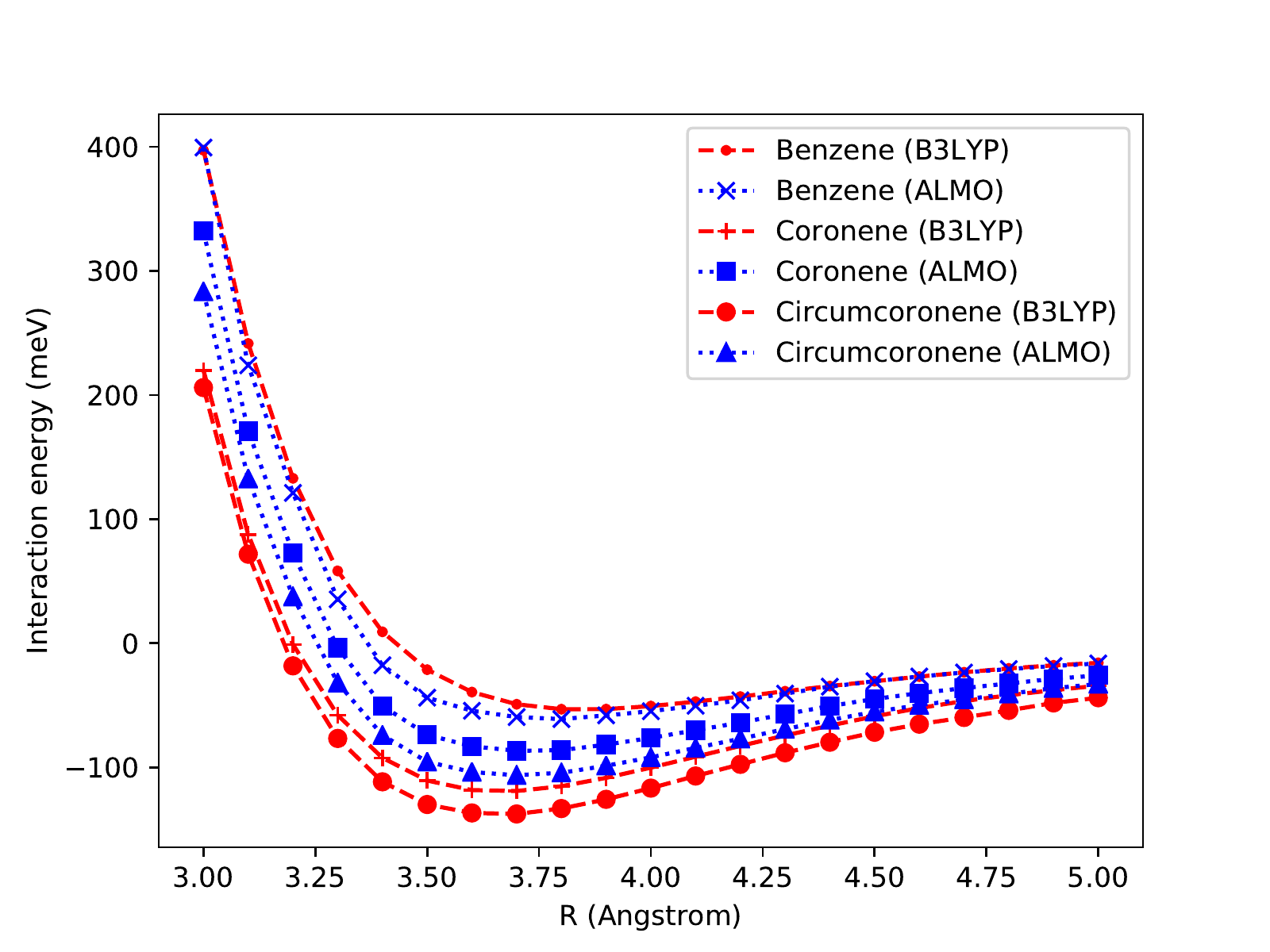} \\
\ce{H2} & \ce{NO2} \\
\includegraphics[width=0.49\textwidth]{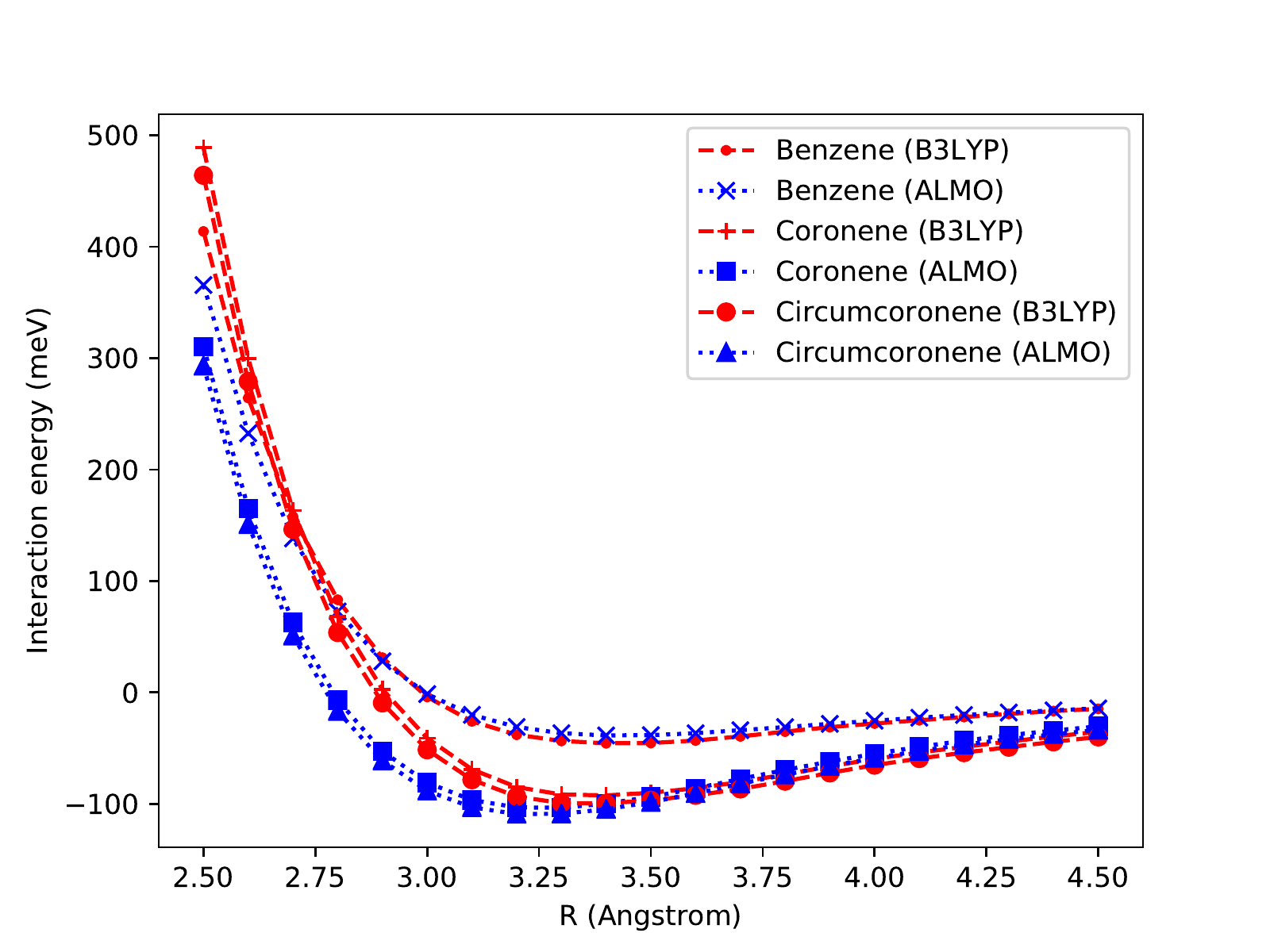} &
\includegraphics[width=0.49\textwidth]{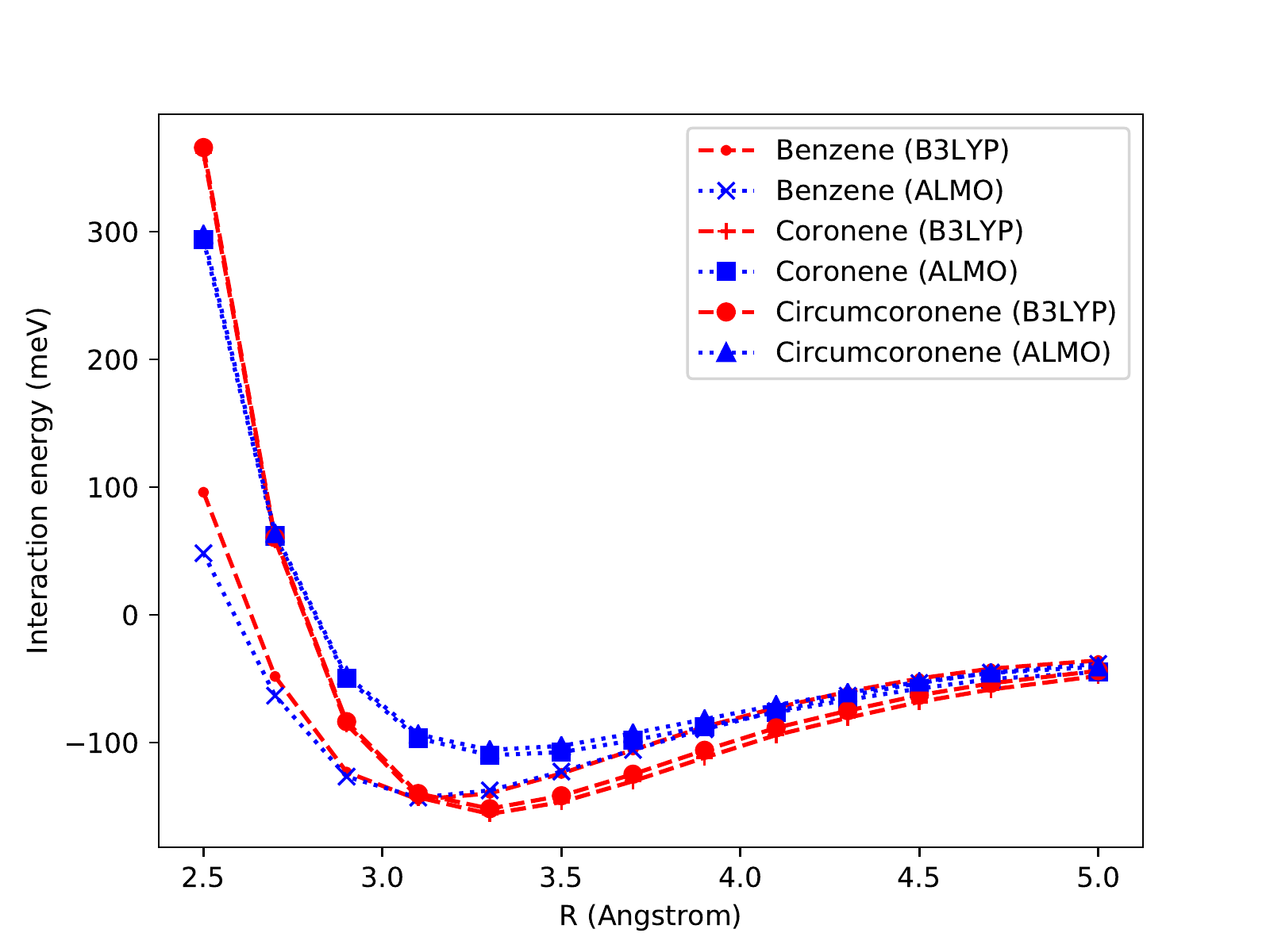}\\
\ce{Ar} & \ce{H2O} 
\end{tabular}

\caption{Interaction energy curves for \ce{H2}, \ce{NO2}, \ce{H2O}, and \ce{Ar} (clockwise from top left) interacting with increasing sizes of carbon flake, comparing the ALMO+CCD(T) results with those of B3LYP+D3, both at the AVTZ level.}
\label{fig.graphcurves}
\end{figure}

To assess our choice of functional for gas-surface interactions with graphene, we now consider complexes of \ce{H2}, \ce{NO2}, \ce{H2O}, and \ce{Ar} with a series of hydrogen-terminated carbon flakes. These flakes are benzene, coronene, and circumcoronene, the latter of which can be seen in Figure~\ref{fig.surfaces}. There are several possible interaction positions for the gas substrate, of which we have selected directly above the central ring of each flake. The reason for this is to minimise any asymmetrical finite size effects. Scans of the distance between the molecule and the surface were again performed for all systems using both B3LYP+D3 and ALMO+rCCD(T), for separations between 2.5 and 5~\AA. All the binding constants, calculated in the same manner as earlier, can be found in Table~\ref{table.specconsts2}, while interaction energy curves are presented in Figure~\ref{fig.graphcurves}.

While the number of carbon atoms, and thus the radial extent, of the flakes increases linearly, the ALMO+rCCD(T) curves show less separation between the two largest flakes than between the two smallest. This is to be expected, as it should tend towards a limiting value as the surface extent increases. Two things are clear, however: convergence is not rapid, with the difference between coronene and circumcoronene well-depths still being around 10~meV for all the molecules except \ce{H2O}; and B3LYP performs progressively worse, overall, as the flake size increases. The first observation suggests that, somewhat unsurprisingly, it is necessary to use much larger surface representations, to accurately study these interactions. This is still relevant in periodic calculations, though, as it implies that a fairly large unit cell is necessary to avoid finite-size effects, given that a circumcoronene flake is roughly equivalent to a 5x5 unit. 

The second observation is more worrying, as it suggests that the errors in dispersion inherent in using empirically-corrected DFT compound as the flake size increases. For all the systems except argon, the discrepancy is so large that the B3LYP coronene curve lies below the circumcoronene ALMO curve, despite those for benzene being in the opposite orientation. This is particularly acute for the hydrogen substrate, where the interaction is purely dispersive, and may be a reflection of the importance of many-body correlation effects that are present in coupled-cluster type methods, but not in DFT or the D3 correction. For \ce{NO2} and \ce{H2O}, where electrostatics are likely to be more important, B3LYP+D3 performs somewhat better, but still with a sizeable discrepancy. The tendency towards overbinding in the energy also corresponds to an underestimation of the equilibrium binding distance, which is perhaps the most experimentally relevant quantity. However, for the fairly minimal cost involved, a discrepancy on the order of 30~meV and 0.15~\AA~may be acceptable in future applications. For the argon atom, B3LYP+D3 generally performs well in the long range, but with a considerable underestimation in the short range. This is consistent with our findings  earlier with respect to noble gas dimers. 

The \ce{H2}, \ce{NO2}, and \ce{Ar} complexes all behave in a qualitatively similar manner with regard to increasing system size. The \ce{H2O} complexes, however, show a considerable decrease in binding as the flake size increases. There is also a substantially smaller difference between the coronene and circumcoronene energies, and a significant shift in the equilibrium binding distance from around 3.2~\AA~to 3.4~\AA. Voloshina and coworkers~\cite{Voloshina2011} found that, for periodic graphene with water in the central position and down orientation (equivalent to ours), a much shorter binding distance of 2.6~\AA, but a similar well depth of 123~meV. This was calculated using an incremental coupled-cluster approach, relying  on periodic local MP2, which is known to over-bind similar systems~\cite{Ma2019}, and with a mixed double/triple-zeta unaugmented basis set for the graphene.  More recently, work by Brandenburg and colleagues~\cite{Brandenburg2019} gave periodic full CCSD(T) and diffusion Monte-Carlo (DMC) results for the same systems (called the 2-leg orientation in their work), again using a mixed zeta basis set. Their results show a similar decrease in magnitude of the well depth going from benzene (136~meV for both L-CCSD(T) and DMC) to periodic graphene (87 and 99~meV for p-CCSD(T) and DMC, respectively).  These are in excellent agreement with our \ce{H2O\bond{-}Circumcoronene} ALMO+rCCD(T) results. They calculated binding distances of 3.3~\AA~(benzene) and 3.4~\AA~(graphene). Interestingly, their RPA results - the theory on which the rCCD correction to ALMO is based - give smaller binding energies (e.g. 82~meV for graphene), which is then improved by adding a GW-type correction. This suggests an interesting avenue of exploration for extending the ALMO+rCCD method to periodic systems. 

\begin{table}
\caption{Binding constants for \ce{H2}, \ce{NO2}, \ce{Ar}, and \ce{H2O} interacting with carbon flakes, ordered by their number of carbon atoms, $n_{\text{C}}$. These are benzene (6), coronene (24), and circumcoronene (54). }
\label{table.specconsts2}

\begin{tabular}{lrrrrrr}
\toprule
& \multicolumn{3}{c}{ALMO+rCCD(T)} & \multicolumn{3}{c}{B3LYP+D3} \\
Molecule / $n_{\text{C}}$ & 6 & 24 & 54 & 6 & 24 & 54 \\
\midrule
 \multicolumn{7}{l}{$R_0$  (\AA)} \\
\ce{NO2} &  3.80 &  3.74 &  3.72 &  3.86 &  3.68 &  3.68 \\
\ce{H2} &   3.03 &  3.00 &  2.97 &  3.02 &  2.97 &  2.92 \\
\ce{Ar} &  3.45 &  3.28 &  3.26 &  3.45 &  3.40 &  3.39 \\
\ce{H2O} &  3.18 &  3.42 &  3.41 &  3.14 &  3.38 &  3.37 \\
\midrule
 \multicolumn{7}{l}{$\epsilon_0$ (meV)} \\
\ce{NO2}  & -60.64 & -87.17 & -106.40 & -54.00 & -119.16 & -137.57 \\
\ce{H2}  & -32.64 & -41.64 & -48.84 & -29.60 & -51.81 & -57.87 \\
\ce{Ar}  & -38.92 & -103.58 & -110.26 & -45.94 & -92.45 & -99.67 \\
\ce{H2O}  & -127.62 & -113.84 & -109.27 & -144.74 & -158.22 & -153.81 \\
\bottomrule
\end{tabular}

\end{table}

\section{Conclusions}

The accurate description of dispersion forces is an important goal in the study of molecule-surface interactions, and presents a difficult challenge for computational methods. In the present investigation, we have considered a number of different classes of dispersion-bound complexes, and assessed empirically corrected density functionals against high-level ab initio methods. These systems have ranged from noble gas dimers to gas molecules on large carbon flakes, with each class presenting different problems to consider. Several of these have been highlighted by our results. 

Firstly, the inclusion of the D3 correction is paramount to give anything resembling reasonable results. The uncorrected functionals almost universally failed to give any binding for many systems, or were in error by at least 100~meV. This was particularly acute for the noble gases, where core-electron correlation effects can be particularly important, and are unlikely to have been accounted for in the design of either the functional or the correction. Less obvious, however, is that there is no significant difference between the functionals, even when moving to those specifically designed for these kinds of system (e.g. M06-2X). On the other hand, however, we see that the approximate coupled-cluster method based on absolutely localised molecular orbitals, ALMO+rCCD(T), gives results within around 10~meV of the full CCSD(T) results in all cases. This continues the promising trend seen in the original paper~\cite{Shaw2019b}, but also demonstrates that the inclusion of triple excitations, here through a perturbative correction, is vital. 

Moreover, we have applied this method for the first time to much larger systems. Worryingly, we find that D3-corrected DFT gives increasingly poor results as the extent of the carbon flake increases. As we are reaching the limit of what is possible with the correlated method, it would thus be prudent to consider alternatives for fully extended systems. An excellent prospect would be GW-based methods, and efforts could be made to extend the ALMO+rCCD method - based on an RPA-like formalism - to follow this approach. Certainly, the investigation of dispersion interactions on surfaces requires a level of accuracy that the density functionals considered here do not seem to provide. It is plausible that double-hybrids would perform better, but there is also no particular computational advantage in using them. Further exploration is essential. 

\begin{acknowledgement}

This work was supported by the Australian Government through the Australian Research Council (ARC) under the Centre of Excellence scheme (project number CE170100026). It was also supported by computational resources provided by the Australian Government through the National Computational Infrastructure National Facility and the Pawsey Supercomputer Centre.

\end{acknowledgement}

\begin{suppinfo}
Optimised geometries, interaction energies, and binding properties for all systems and methods. 
\end{suppinfo}

\bibliography{disprefs.bib}

\end{document}